\documentclass[a4]{aa} 

\newcommand{\gtsim}{\protect\raisebox{-0.5ex}{$\:\stackrel{\textstyle >}
        {\sim}\:$}}
\newcommand{\ltsim}{\protect\raisebox{-0.5ex}{$\:\stackrel{\textstyle <}
        {\sim}\:$}}

\usepackage{graphicx}
\usepackage[authoryear]{natbib}
\bibpunct{(}{)}{;}{a}{}{,} 
 
\begin{document}

\title{Spots, plages, and flares on $\lambda$~Andromedae and II~Pegasi\thanks{Based on observations collected 
at Catania Astrophysical Observatory (Italy) and Ege University Observatory (\.Izmir, Turkey).}}
   
\author{A. Frasca  \inst{1}  \and K. Biazzo \inst{1}  \and G. Ta\c{s} \inst{2} \and S. Evren \inst{2} \and A. C. Lanzafame \inst{3}}
\offprints{A. Frasca}
\mail{afrasca@oact.inaf.it}

  \institute{INAF - Catania Astrophysical Observatory, via S. Sofia 78, I--95123 Catania, Italy
 \and Ege University Observatory, Bornova, \.{I}zmir, Turkey 
 \and Department of Physics and Astronomy, Astrophysics Section, University of Catania, via 
 S. Sofia 78, I--95123  Catania, Italy}

\date{Received 21 May 2007 / Accepted 7 November 2007}

\abstract
{}
{We present the results of a contemporaneous photometric and spectroscopic monitoring of two RS~CVn 
binaries, namely \object{$\lambda$~And} and \object{II~Peg}. The aim of this work is to investigate the behavior of surface 
inhomogeneities in the atmospheres of the active components of these systems which have nearly the same temperature but
different gravity.} 
{The light curves and the modulation of the surface temperature, as 
recovered from line-depth ratios (LDRs), are used to map the photospheric spots, while the H$\alpha$ emission 
has been used as an indicator of chromospheric inhomogeneities.} 
{The spot temperatures and sizes were derived from a spot model applied to the contemporaneous light and temperature 
curves. We find larger and cooler spots on II~Peg ($T_{\rm sp} \simeq 3600$\,K) compared to $\lambda$~And 
($T_{\rm sp} \simeq 3900$\,K); this could be the result of both the different gravity and the higher activity level of 
the former.
Moreover, we find a clear anti-correlation between the H$\alpha$ emission and the photospheric diagnostics 
(temperature and light curves).
We have also detected a modulation of the intensity of the \ion{He}{i} D$_3$ 
line with the star rotation, suggesting the presence of surface features also in the upper chromosphere of these stars.
A rough reconstruction of the 3D structure of their atmospheres 
has been also performed by applying a spot/plage model to the light and temperature curves and to the 
 H$\alpha$ flux modulation. In addition, a strong flare affecting the H$\alpha$, the \ion{He}{i} D$_3$, and the
 cores of \ion{Na}{i} D$_{1,2}$ lines has been observed on II~Peg.}
{The spot/plage configuration has been reconstructed in the visible component of $\lambda$~And and II~Peg which have
nearly the same temperature but very different gravity and rotation periods. A close spatial association of photospheric 
and chromospheric active regions, at the time of our observations, has been found in both stars. 
Larger and cooler spots have been found on II~Peg, the system with the active component of higher gravity and higher 
activity level.   The area ratio of plages to spots seems to decrease when the spots 
get bigger. Moreover, with the present and literature data, a correlation between the temperature difference 
$\Delta T = T_{\rm ph}-T_{\rm sp}$ and  the surface gravity has been also suggested.}

   
\titlerunning{Spots, plages, and flares in $\lambda$~And and II~Peg.}
\authorrunning{A. Frasca et al.}
\maketitle

\section{Introduction}

\label{sec:Intro}
The rotational modulation of brightness and other photospheric diagnostics in late-type stars is commonly 
attributed to starspots on their photospheres. 
The presence of chromospheric inhomogeneities similar to the solar plages is pointed out by several
studies based on the H$\alpha$ and/or other diagnostics \citep[see, e.g., ][]{Stra93a, Cata96, Cata00, Fra98, Fra00a}. 

The simultaneous study of photospheric (spots) and chromospheric (plages) active regions (ARs) is important 
for a better understanding of the physical processes occurring during the emersion of magnetic flux tubes 
from below the photosphere. Indeed, the relative position and size of ARs at different atmospheric levels 
can provide information on the magnetic field topology. In the past ten years, rotational modulation 
due to surface inhomogeneities at photospheric and chromospheric level has been revealed in several RS~CVn 
binaries, \citep{Fra98, Cata00, Bia06}. 

Simultaneous photometric and spectroscopic observations have frequently shown a spatial association 
between spots and plages in RS~CVn systems  \citep{Rodo87,Cata96,Fra98} as well as in young mildly-active 
solar-type stars \citep{Fra00a,Bia07}. A  close plage-spot association has been also detected in very active main-sequence 
single stars, like the rapidly rotating star LQ~Hya \citep{Stra93a}.
Also in close binary systems \citep[e.g. TZ~CrB,][]{Fra97} and in extremely active stars, like the 
components of the contact binary VW~Cep \citep{Fra96}, there are evidences of spot-plage associations. 
Moreover, for some RS~CVn binary systems there is some indication of a systematic longitude lag of 30$^\circ$--50$^\circ$ 
between the plages and spots \citep{Cata96,Cata00}.

In recent works \citep[][ hereafter Paper I and Paper II, respectively]{Cata02,Fra05}, we 
presented the results of a contemporaneous spectroscopic and photometric monitoring of three 
active single-lined (SB1) RS~CVn binaries (\object{VY~Ari}, \object{IM~Peg}, and \object{HK~Lac}), showing that it is 
possible to recover fairly accurate spot temperature and size values by applying a spot model to contemporaneous 
line-depth ratios (LDRs) variation and light curves. 
Subsequently we have investigated the active region topology in these stars at both chromospheric and photospheric 
layers \citep[][ hereafter Paper III]{Bia06}. 

In the present paper, a similar analysis is applied to two other active SB1 RS~CVn's, namely $\lambda$~And and II~Peg,
which are stars with similar effective temperatures ($T_{\rm eff}\simeq$\,4700 and 4600\,K, respectively) but with  
 different gravities,  being $\log g\simeq$\,2.5 for $\lambda$~And \citep{Dona95} and $\log g\simeq$\,3.2
 for II~Peg \citep{Berdy98b}.

$\lambda$~And (HD~222107) is a bright ($V=3\fm82$) and active giant, classified as G8 IV-III. \citet{Calder38} was the first 
to discover its photometric variability, whose amplitude sometimes reaches 0$\fm$30. Six years later, 
\citet{Walker44} 
showed that $\lambda$~And is a SB1 with an almost circular orbit of period 20$\fd$5212. It is an atypical member of 
the RS~CVn class, because it is largely out of synchronism, its rotational period being 53$\fd$952 \citep{Stra93b}.
Indeed, in most RS~CVn binaries the rotational period of both components is equal to the orbital period of the 
system within a few percent. 
Thus $\lambda$~And, whose orbit is very close to be circular, is a puzzle for the theory of tidal friction \citep{zahn77}, 
which predicts that rotational synchronization in close binaries should precede orbit circularization. It is one of the 
brightest of all known chromospherically active binaries. Photoelectric light curves of $\lambda$~And 
\citep{BoppNoah80,PoeEaton85} show that its photometric period is somewhat variable, probably due to differential 
rotation and latitude drift of the spots during magnetic cycles. In addition, there is a long-term cycle of about 11 yr 
in the mean brightness. Photometric long-term studies have been performed also by \citet{Henry95}.  
Moreover, the H$\alpha$ emission has been found to be rotationally modulated and anti-correlated with the light curve 
\citep{BalDup82}.

II~Peg (HD 224085) has been classified as a K2-3 IV-V SB1 by \citet{Ruci77} who noticed that its 
photometric period was quite close to 6$\fp$724183, as determined by \citet{Halli52}  
for its essentially circular orbit. It was classified as an RS~CVn system by \citet{Vogt81a,Vogt81b}   
who found an amplitude of the $V$ light curve of 0$\fm$43. 
II~Peg is among the most active RS~CVn binaries and belongs to a small subset of binaries, 
including V711~Tau and UX~Ari, in which H$\alpha$ always appears in emission \citep{NatRam81}. 
The first photoelectric light curves have been published by \citet{Chuga76}  
Long-term studies \citep[e.g.,][]{Henry95, Rodo00} have shown dramatic changes of the photometric wave, from almost 
sinusoidal, to irregular or flat. Moreover, \citet{Dona97} have clearly detected the magnetic field on this star. 
The so called flip-flop phenomenon (in which the dominant part of the spot activity changes the longitude every few 
years) has been reported in II~Peg by, e.g., \citet{BerdyTuo98} and \citet{Rodo00} and theoretically analyzed by 
\citet{ElsKor05}.

The aim of the present work is to investigate the starspot characteristics of this two active stars with the same
temperature but different gravity and activity level, as well as to study the location of the excess H$\alpha$ 
emission and the degree of spatial superposition between surface inhomogeneities at different atmospheric levels.

\section{Observations and reduction}

 \label{sec:Obs}

\subsection{Spectroscopy}
Spectroscopic observations have been performed in 1999 and 2000 at the {\it M. G. Fracastoro station} (Serra La Nave, 
Mt. Etna) of Catania Astrophysical Observatory with FRESCO (Fiber-optic Reosc Echelle Spectrograph of Catania 
Observatory), the \'echelle spectrograph connected to the 91-cm telescope through an optical fiber with a 200-$\mu$m
core diameter. The spectral resolution was $R=\lambda/\Delta\lambda\,\simeq\,$14\,000, with a 2.6-pixel sampling.

The data reduction was performed by using the {\sc echelle} task of the IRAF\footnote{IRAF is distributed by the 
National Optical Astronomy Observatory, which is operated by the Association of the Universities for Research in 
Astronomy, inc. (AURA) under cooperative agreement with the National Science Foundation.} package following the 
standard steps of  background subtraction, division by a flat field spectrum given by a halogen lamp, wavelength 
calibration using the emission lines of a Thorium--Argon lamp, and normalization to the continuum through a 
polynomial fit. 

 We have removed the telluric water vapor lines at the H$\alpha$ wavelengths
using the spectra of \object{Altair} (A7~V, $v\sin i\,=\,245$~km\,s$^{-1}$) acquired during the 
observing runs. These spectra have been normalized, also inside the very broad 
H$\alpha$ profile, to provide valuable templates for the water vapor lines.
An interactive procedure, allowing the intensity of the template 
lines to vary (leaving the line ratios unchanged) until a satisfactory 
agreement with each observed spectrum is reached, has been applied to 
correct the observed spectra for telluric absorption \citep[see ][]{Fra00a}.

\subsection{Photometry}
The photometric observations were performed in the $B$, $V$, and $R$ Johnson filters 
at the Ege University Observatory.
The observations were made with  an un-refrigerated Hamamatsu R4457 photometer attached to the 48-cm 
Cassegrain telescope. 

$\lambda$~And was observed from July 9  to November 2, 1999 for a total of 35 nights, using 
$\psi$~And and $\kappa$~And as comparison ($C$) and check ($Ck$) star, respectively.
II~Peg was observed from July 3 to November 21, 2000 for a total of 25 nights, using HD~224083 and
BD+27$\degr$4648 as comparison and check star, respectively.

The differential magnitudes, in the sense of variable minus comparison ($V-C$), were corrected for
atmospheric extinction using the seasonal average coefficients for the Ege University Observatory.
The light curves were obtained by averaging  individual data points  taken in the same night
(from 5 to 20).
 The standard deviation of each observed point, as measured from the differential magnitude $V-C$ and
 $C-Ck$, ranges from $\pm0\fm005$ to $\pm0\fm015$.

\section{Data analysis}

\subsection{H$\alpha$ line analysis}

The hydrogen H$\alpha$ line is one of the most useful and easily accessible indicators of chromospheric activity 
in the optical spectrum. Furthermore, it is very effective, both in the Sun and in active stars, for detecting 
chromospheric plages, due to their high contrast against the surrounding chromosphere.

II~Peg is one of the few RS~CVn stars which displays the H$\alpha$ line always in emission above the local continuum.
For the majority of the active binaries, instead, only a filling-in of the H$\alpha$ core can be seen. In these cases,
the ``spectral synthesis" method, based on the comparison with synthetic spectra from radiative equilibrium models 
or observed spectra of non-active standard stars (reference spectra), has been successfully used 
\citep[see, e.g.,][]{Herb85, Bar85, Fra94, Montes95}.
The difference between observed and reference spectrum provides, as residual, the net chromospheric H$\alpha$ emission, 
which can be integrated to find the total radiative losses in the line.

Since II~Peg and $\lambda$~And are SB1 systems, we used only one standard star spectrum to reproduce their observed 
spectra. This standard star spectrum has been rotationally broadened by convolution with a rotational profile 
with $v\sin i=22.6$ km\,s$^{-1}$ for II~Peg and $v\sin i=6.5$ km\,s$^{-1}$ for $\lambda$~And to mimic the active star 
in absence of chromospheric activity. Figures~\ref{fig:lam_spettri1999} and \ref{fig:ii_spettri2000} show samples of 
spectra in the H$\alpha$ region of $\lambda$~And and II~Peg, respectively. The residual H$\alpha$ equivalent width, 
$\Delta EW_{\rm H\alpha}$, has been measured by integrating all the emission profile in the difference spectrum 
(see the lower panel of Fig.~\ref{fig:lam_spettri1999}).

The error, $\sigma(\Delta EW_{\rm H\alpha})$, has been evaluated by multiplying the integration range by the photometric 
error on each point. This latter has been estimated by the standard deviation of the observed flux values on the 
difference spectra in two spectral regions near the H$\alpha$ line. Although the errors $\sigma(\Delta EW_{\rm H\alpha})$ 
are generally a few hundredths of an \AA, a very small systematic error in $\Delta EW_{\rm H\alpha}$, due to a possible 
contamination by chromospheric emission in the H$\alpha$ core of the reference stars, could be still present. 
However, this contribution can be completely neglected for very active stars, like the RS~CVn systems here investigated. More information about the 
spectral synthesis method can be found in \citet{Fra94}.

\subsection{The Helium D$_{3}$ line}

As a further indicator of chromospheric activity we have analyzed also the \ion{He}{i} $\lambda$\,5876 (D$_{3}$) line, 
which is normally seen as an absorption feature in active stars \citep[e.g., ][]{Huene86,Bia06,Bia07}.
The reference spectra adopted do not show any \ion{He}{i} absorption nor emission, as expected for non-active
stars. In this case, the spectral subtraction technique allows to emphasize the \ion{He}{i} line, cleaning the
spectrum from nearby photospheric absorption lines, and to measure its equivalent width, $EW_{\rm HeI}$.
For  $EW_{\rm HeI}$ we have adopted the usual convention that an absorption line has a positive $EW$.

The helium $\lambda$\,5876 line, because of its high excitation potential, is a good tracer of  
the regions of higher temperature and excitation in solar and stellar chromospheres. 
Recent models seem to indicate that the primary mechanism responsible for the formation of the \ion{He}{i} triplet is the 
collisional excitation and ionization by electron impact followed by recombination cascade \citep{Lan95}.

In the Sun, the \ion{He}{i} D$_{3}$ line appears as an absorption feature in plages and weak flares and in 
emission in strong flares \citep[e.g., ][]{Ziri88}.
In active stars, the \ion{He}{i} D$_{3}$ is usually observed in absorption and sometimes in emission, like during
flare events in the most active RS~CVn stars \citep{Montes97, Montes99, Garcia03}.
This is related to the electronic temperature and density in the emitting region \citep[e.g., ][]{Ziri88,Lan95}.
A contribution to the emissivity from overionisation due to coronal EUV back-radiation is expected to play a 
role when the transition region pressure is below 1 dyne\,cm$^{-2}$ \citep{Lan95}.
This contribution is therefore relevant for moderately active stars. In any case, our analysis in independent on
the detailed mechanism of formation.

\subsection{Photospheric temperature from LDRs}

Line-depth ratios (LDRs) can be used for detecting the temperature rotational modulation in active RS~CVn stars. 
Such diagnostics allow to detect temperature variations as small as 10--20\,K at the resolution of our 
spectra and with a good signal-to-noise ratio (S/N $\geq 100$). The precision of this 
method is improved by averaging the results from several line pairs, as discussed in Papers~I and II. 
Typically we used from seven to fifteen line pairs, depending on the star's $v\sin i$, in the wavelength region 
around 6250~\AA ~to produce an average value of the star photospheric temperature at each phase. 

We found a clear rotational modulation of the photospheric temperature recovered from LDRs for
three SB1 RS~CVn binaries, namely VY~Ari, IM~Peg, and HK~Lac (Paper~I). 
In Paper~II we showed that these temperature curves are very well correlated with the contemporaneous light curves. 
Moreover, the simultaneous modeling of the temperature and light curves enabled us to derive temperature and size 
of the starspots.

In principle, rotational line-broadening must be taken into account in a technique such as the LDR. However,
$v\sin i$ is usually smaller than the instrumental resolution and rotational broadening can be safely ignored.
This is not the case for II~Peg, whose rotational broadening ($v\sin i=22.6$~km\,s$^{-1}$) is larger than the 
resolution of our spectra. In this case we have therefore produced a LDR--$T_{\rm eff}$ calibration using 
rotationally broadened reference spectra.
The values of temperature deduced for II~Peg with these calibrations are, however, close to those found with the  
calibration at $v\sin i=0$ reported in Paper~I.
 
Further information about the LDR method can be found in Papers~I and II.

\section{Results}

\subsection{$\lambda$ And}

\begin{figure}
\centering
\includegraphics[width=7cm]{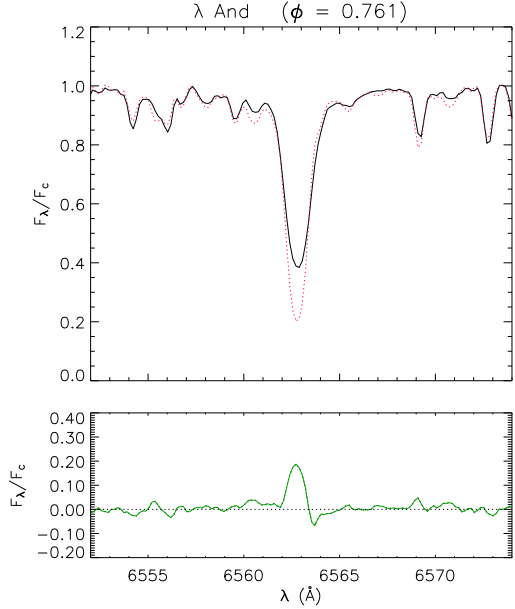}	
\caption{{\it Top panel}:  Example of observed spectrum of $\lambda$~And at the rotational phase $\phi=0\fp761$ 
in the H$\alpha$ region (full line) together with the non-active template (dotted line). {\it Lower panel}: 
Difference between observed and template spectra emphasizing the core chromospheric emission.
}
\label{fig:lam_spettri1999}
\end{figure}

\begin{figure}
\hspace{-1.7cm}
\includegraphics[width=12.7cm]{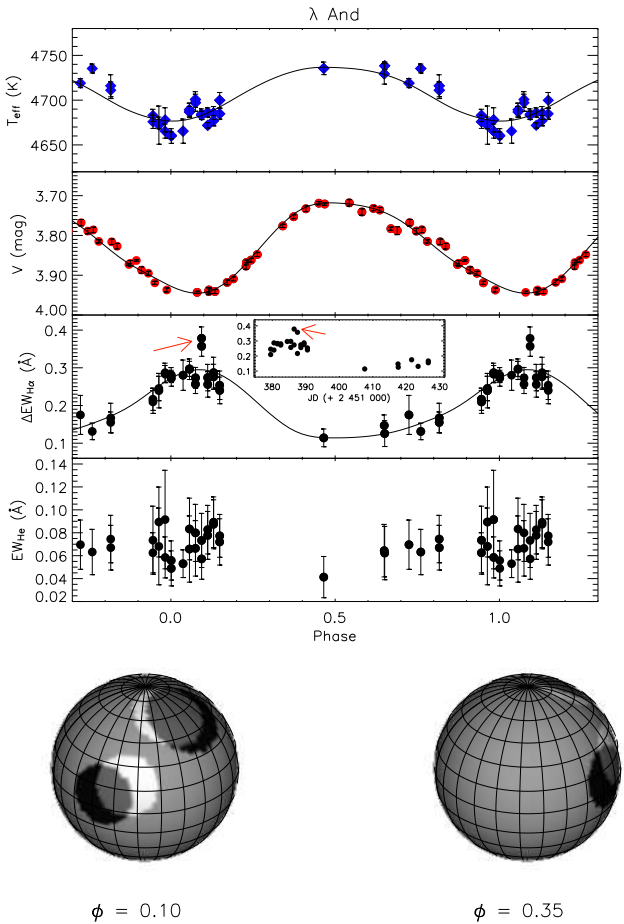}	
\caption{{\it From top to bottom.} Johnson $V$ photometry, 
$T_{\rm eff}$, $\Delta EW_{\rm H\alpha}$, and $EW_{\rm HeI}$ as a function of the rotational phase for $\lambda$~And.
The synthetic temperature, light, and $EW_{\rm H\alpha}$ curves, calculated with our spot/plage model 
(see Sect.~\ref{sec:spot}), are reproduced with full lines in each box. The inset displays the $\Delta EW_{\rm H\alpha}$
versus Julian Day. A schematic map of the spot and plage distributions, as seen at two different rotational phases, 
is also shown in the bottom.} 
\label{fig:lambda_and_tm_v_ew}
\end{figure}

For $\lambda$~And, nine LDRs have been used and transformed in temperature variations by means of the LDR--$T_{\rm eff}$ 
calibration with no rotational broadening. The rotational velocity of $\lambda$~And, namely 6.5 km\,s$^{-1}$ \citep{Dona95}, 
is in fact lower than the FRESCO resolution of about 7 km\,s$^{-1}$, 
so that no correction for rotational broadening is needed. 
All the LDRs converted into temperature and combined in a single temperature curve
lead to a fairly well-defined temperature variation as a function of the rotational phase and well correlated with the 
optical light curve (Fig.~\ref{fig:lambda_and_tm_v_ew}). Unfortunately, the very long rotation period has prevented us 
to obtain a complete phase coverage, but we have enough data around the maximum and minimum of the curve for 
performing a meaningful analysis.

The rotational phases have been derived according to the following ephemeris
\begin{equation}
HJD_{\phi=0} = 2\,443\,829.2+53\fd952\,\times\,E\,,
\label{Eq:ephem_lambda}
\end{equation}
{\noindent where the initial heliocentric Julian day is that of \citet{Henry95} and the rotational period is
taken from \citet{Stra93b}. 
The temperature maximum, with a value of 4740 K, occurs at phase 
$\phi\simeq0\fp65$. The full amplitude variation of the effective temperature is $\Delta T_{\rm eff}\simeq 80$\,K, 
corresponding to about 2\% of 
the average temperature. The light curve displays a variation amplitude $\Delta V=0\fm$225.}

$\mu$~Peg (G8\,III), with $B-V=0\fm934$, has been used as non-active H$\alpha$ reference, because it is of the same 
spectral type and nearly the same color as $\lambda$~And. The H$\alpha$ line of $\lambda$~And is always in absorption 
but with a variable filling-in in the core (see Fig.~\ref{fig:lam_spettri1999} for an example). 
The values of the residual emission $\Delta EW_{\rm H\alpha}$ with their errors are plotted in 
Fig.~\ref{fig:lambda_and_tm_v_ew} as a function of the rotational phase. Two $\Delta EW_{\rm H\alpha}$ values taken in 
the same nights (marked with an arrow) could be related to a mild flare event. 
In these hypothesis, we can only give an upper limit of two days for the flare duration, because there is no sign of 
H$\alpha$ enhancements either in the night before or in that following the event.
It is also very difficult to estimate the flare energetics because we don't know the flare duration and the H$\alpha$
level at the flare peak.
However, we have roughly evaluated the energy released in the H$\alpha$ line during the event assuming a peak value 
$\Delta EW_{\rm H\alpha}=0.37$\,\AA ~(underestimate) and a flare duration of 2 days (overestimate).
We have converted the net H$\alpha$ equivalent into luminosity by means of the equation:
\begin{eqnarray}
L_{\rm H\alpha} & = & L_{6563}\Delta EW_{\rm H\alpha} = 4\pi d^2 f_{6563}\Delta EW_{\rm H\alpha} \nonumber\\
		& = & 4\pi d^2 \frac{F_{6563}}{F_{5556}}10^{(-0.4V_0-8.451)}\Delta EW_{\rm H\alpha},
\label{eq:flare}
\end{eqnarray}
{\noindent where $L_{6563}$ and $f_{6563}$ are the luminosity and the flux at Earth of the continuum at 
$\lambda=6563$ \AA, $d$ is the distance and $10^{(-0.4V_0-8.451)}$ is the Earth flux at $\lambda=6563$ \AA~from 
a star of $V_0$ (de-reddened) magnitude. The continuum flux-ratio $\frac{F_{6563}}{F_{5556}}$ has been evaluated 
using NextGen synthetic low-resolution spectra \citep{Haus99}. 
We find a luminosity at peak 
$L_{\rm H\alpha}\ltsim 3\times10^{30}$ erg\,s$^{-1}$ and an approximate value for the total energy emitted in the 
H$\alpha$ line of $E_{\rm H\alpha}\approx 8\times10^{34}$ erg.}

Besides this possible flare event, a clear anti-correlation between 
temperature and H$\alpha$ emission is apparent with a similar shape of the curves. \ion{He}{i} D$_3$ 
absorption has been also detected, but no clear modulation emerges from the scatter of the data. 

The values of $T_{\rm eff}$, $\Delta EW_{\rm H\alpha}$, and $EW_{\rm HeI}$  for all the observed spectra are listed in 
Table~\ref{tab:halpha_lambda}.

\begin{table}[t]   
\caption{Temperature values and parameters of the subtracted spectra of $\lambda$~And.}
\label{tab:halpha_lambda}
\scriptsize
\begin{center}
\begin{tabular}{ccccc}
\hline
\hline
HJD & Phase & $T_{\rm eff}$ & $\Delta EW_{\rm H\alpha}$ & $EW_{\rm HeI}$\\ 
(+2\,400\,000) & & (K) & (\AA) & (\AA) \\ 
\hline
51\,379.527 & 0.945 & 4683$\pm$ 7& 0.22$\pm$0.03& $0.063\pm$0.018\\
51\,379.535 & 0.945 & 4676$\pm$ 7& 0.21$\pm$0.03& $0.074\pm$0.030\\
51\,380.500 & 0.963 & 4675$\pm$ 6& 0.25$\pm$0.03& $0.068\pm$0.034\\
51\,380.512 & 0.964 & 4672$\pm$21& 0.24$\pm$0.05& $0.089\pm$0.031\\
51\,381.527 & 0.982 & 4665$\pm$ 8& 0.29$\pm$0.03& $0.092\pm$0.043\\
51\,381.539 & 0.983 & 4678$\pm$13& 0.28$\pm$0.02& $0.059\pm$0.018\\
51\,382.531 & 0.001 & 4661$\pm$ 4& 0.28$\pm$0.02& $0.056\pm$0.017\\
51\,382.539 & 0.001 & 4660$\pm$ 8& 0.27$\pm$0.02& $0.049\pm$0.015\\
51\,384.480 & 0.037 & 4665$\pm$14& 0.28$\pm$0.04& $0.053\pm$0.012\\
51\,385.527 & 0.056 & 4689$\pm$ 7& 0.30$\pm$0.03& $0.066\pm$0.029\\
51\,385.535 & 0.057 & 4687$\pm$ 6& 0.30$\pm$0.02& $0.083\pm$0.027\\
51\,386.512 & 0.075 & 4701$\pm$ 9& 0.27$\pm$0.02& $0.066\pm$0.024\\
51\,386.520 & 0.075 & 4697$\pm$10& 0.27$\pm$0.03& $0.080\pm$0.025\\
51\,387.520 & 0.093 & 4684$\pm$ 6& 0.38$\pm$0.03& $0.073\pm$0.024\\
51\,387.527 & 0.094 & 4683$\pm$ 5& 0.36$\pm$0.03& $0.057\pm$0.018\\
51\,388.523 & 0.112 & 4672$\pm$ 3& 0.27$\pm$0.04& $0.078\pm$0.021\\
51\,388.531 & 0.112 & 4686$\pm$ 5& 0.26$\pm$0.03& $0.083\pm$0.021\\
51\,389.488 & 0.130 & 4686$\pm$ 7& 0.28$\pm$0.03& $0.087\pm$0.021\\
51\,389.496 & 0.130 & 4677$\pm$ 6& 0.29$\pm$0.04& $0.089\pm$0.022\\
51\,390.488 & 0.148 & 4685$\pm$ 6& 0.25$\pm$0.04& $0.072\pm$0.020\\
51\,390.500 & 0.149 & 4700$\pm$ 9& 0.24$\pm$0.06& $0.077\pm$0.019\\
51\,407.586 & 0.465 & 4736$\pm$ 7& 0.11$\pm$0.02& $0.041\pm$0.018\\
51\,417.520 & 0.650 & 4729$\pm$11& 0.15$\pm$0.03& $0.064\pm$0.023\\
51\,417.590 & 0.651 & 4738$\pm$ 4& 0.13$\pm$0.03& $0.062\pm$0.023\\
51\,421.578 & 0.725 & 4719$\pm$ 5& 0.18$\pm$0.05& $0.070\pm$0.022\\
51\,423.520 & 0.761 & 4735$\pm$ 5& 0.13$\pm$0.02& $0.063\pm$0.020\\
51\,426.547 & 0.817 & 4716$\pm$12& 0.16$\pm$0.03& $0.075\pm$0.021\\
51\,426.555 & 0.817 & 4711$\pm$ 9& 0.17$\pm$0.04& $0.067\pm$0.020\\
\hline
\hline
\end{tabular}
\end{center}
\end{table}
\normalsize

\subsection{II~Pegasi}

\begin{figure}
\centering
\includegraphics[width=9cm,height=11cm]{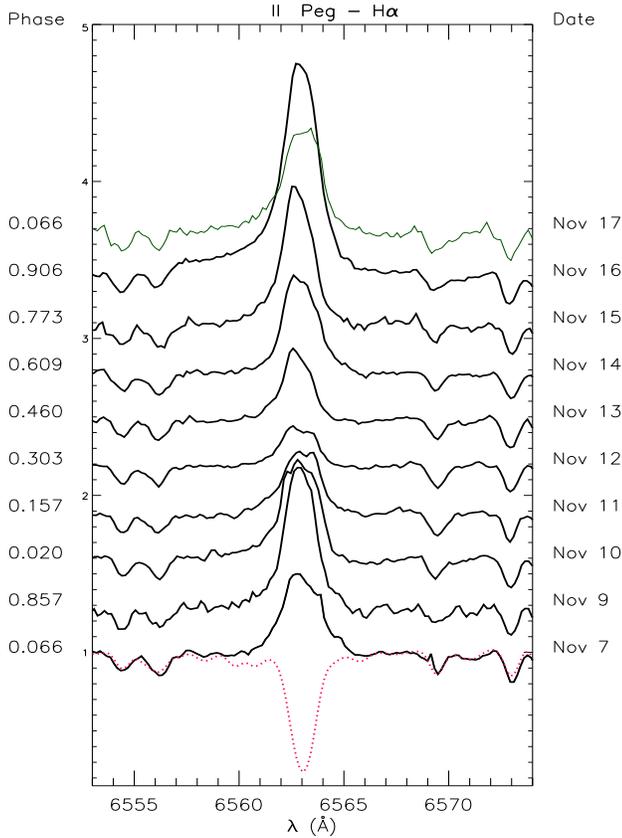}	
\caption[]{Time series of the H$\alpha$ profile of II~Peg (full lines) together with the non-active template (dotted line). 
Note the very strong H$\alpha$ profile with broad wings observed during a strong flare on November 16$^{\rm th}$ followed by
a nearly ``normal" emission profile the night after (thin line).}
\label{fig:ii_spettri2000}
\end{figure}

\begin{figure}[t]
\hspace{-1.7cm}
\includegraphics[width=12.7cm]{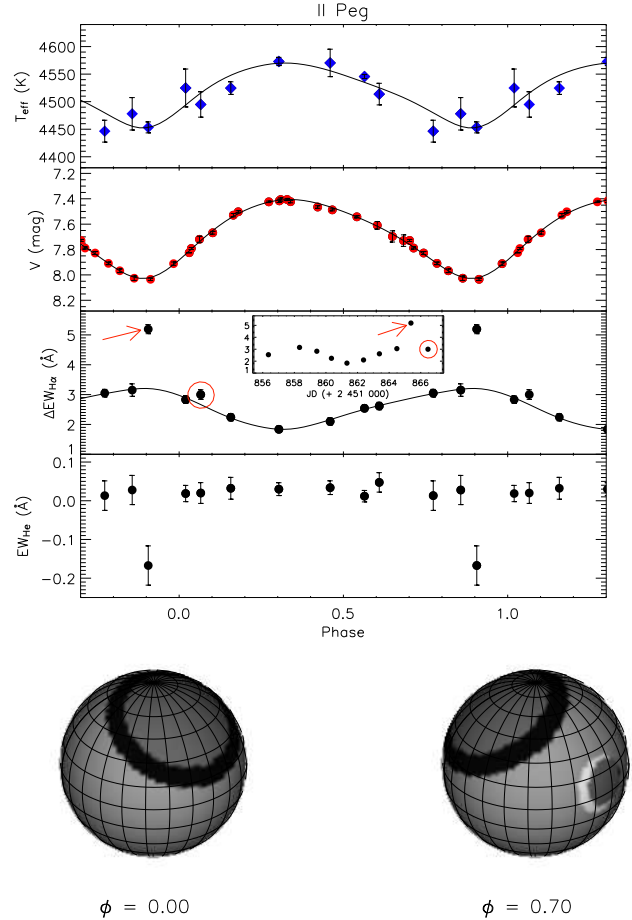}	
\caption[]{{\it From top to bottom.} Johnson $V$ magnitude, averaged $T_{\rm eff}$, total H$\alpha$ emission and 
\ion{He}{i} D$_{3}$ equivalent width as a function of the rotational phase for II~Peg.
The synthetic temperature, light, and $EW_{\rm H\alpha}$ curves, calculated with our spot/plage model 
(see Sect.~\ref{sec:spot}), are reproduced with full lines in each box. The inset displays the $\Delta EW_{\rm H\alpha}$
versus Julian Day with an arrow marking the flare peak. A schematic map of the spot and plage distributions, as seen at 
two different rotational phases, is also shown in the bottom.} 
\label{fig:iipeg_tm_v_ew}
\end{figure}

Seven line-depth ratios have been used for II~Peg. A few line pairs have been disregarded due to severe 
blending at the $v\sin i$ of II~Peg. The LDR variations have been transformed into temperature 
variations by means of the calibration based on standard stars rotationally broadened at the same $v\sin i$ of 
II~Peg, namely 22.6 km\,s$^{-1}$ \citep{Berdy98a}.  The rotational modulation of the temperature and 
$V$ magnitude is shown in the two upper boxes of Fig.~\ref{fig:iipeg_tm_v_ew} as a function of the 
phase calculated according to the following ephemeris 
\begin{equation}
HJD_{\phi=0} = 2\,443\,030.24+6\fd724333\,\times\,E\,,
\label{Eq:ephem_II}
\end{equation}
{\noindent where the rotational period and the initial epoch are taken from \citet{Berdy98a} 
and \citet{HallHenry83},  
respectively. The temperature variation is about 3\%, with an amplitude $\Delta T_{\rm eff}\simeq 130$ K, 
while the light curve amplitude is $\Delta V\simeq0\fm63$. The two curves seem to have a slightly different shape 
and a small shift of the phase of minimum.}

\begin{figure}
\centering
\vspace{-.5cm}
\includegraphics[width=9cm,height=7cm]{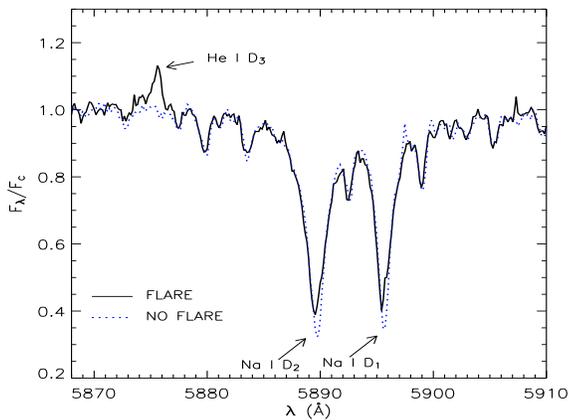}	
\vspace{-.5cm}
\caption[]{Two observed spectra of II~Peg in the region of \ion{He}{i} D$_{3}$ and \ion{Na}{i} D$_{1}$,D$_{2}$
at the flare peak (full line) and in a quiescent phase (dotted line).
Note the strong \ion{He}{i} emission and the remarkable filling-in of the \ion{Na}{i} doublet during the flare.
} 
\label{fig:iipeg_spettro_he_emiss}
\end{figure}

$\delta$~Eri (K0\,IV) has been used as a reference inactive star, whose spectrum has been broadened at the rotational 
velocity of II~Peg. 
A sample of II~Peg spectra in the H$\alpha$ region at different phases is shown in 
Fig.~\ref{fig:ii_spettri2000}. The H$\alpha$ line is always in emission above the continuum level with a small core 
reversal from $\phi=0\fp157$ to $0\fp304$, while a pure emission with asymmetric shape is observed at other phases.
A very strong and broad emission is observed at $\phi=0\fp906$. 
The $\Delta EW_{\rm H\alpha}$ and $EW_{\rm HeI}$ as a function of the rotational phase are plotted 
in Fig.~\ref{fig:iipeg_tm_v_ew} and listed in Table~\ref{tab:halpha_ii} together with the temperature values. 
In Fig.~\ref{fig:iipeg_tm_v_ew}, a sudden increase of the H$\alpha$ equivalent width at HJD=2\,451\,865 ($\phi=0\fp906$), 
which corresponds to the onset of a flare as witnessed also by the H$\alpha$ profile, is marked with an arrow. 
This behavior is also present in the equivalent width of the \ion{He}{i} line, which at that epoch was in emission 
(Fig.~\ref{fig:iipeg_spettro_he_emiss}), unlike outside-flare spectra in which it is always in absorption. 
The strength of the event is also witnessed by the remarkable filling-in of the sodium D$_1$ and D$_2$ lines 
(Fig.~\ref{fig:iipeg_spettro_he_emiss}), similarly to what found for HR~1099 by \citet{Garcia03}.
Since II~Peg was in a quiescent phase on November 15$^{\rm th}$, 2000, it is possible to estimate a lower limit of 
two days for the life of the flare, assuming that the outburst was composed by only one event. 
In fact, one day after the flare peak (November 17$^{\rm th}$, HJD=2\,451\,866) there is still some H$\alpha$ emission 
excess, with respect to the average modulation curve (big circle in Fig.~\ref{fig:iipeg_tm_v_ew}). 
Unfortunately, our photometric data does not include the $U$-band, sensitive to flares, and therefore no comparison 
with photometry can be made.

We evaluated the energy released in H$\alpha$ according to Eq.~\ref{eq:flare}.
 Integrating the ``excess luminosity'' on all the duration 
of the flare ($\gtsim 2$ days), the total energy emitted in the H$\alpha$ line is $E_{\rm H\alpha}^{\rm tot}\gtsim10^{35}$ 
erg. It is worth noticing that this flare has occurred near the $V$ minimum, which denotes a possible spatial association 
with the photospheric spots, in analogy with the most energetic solar flares, the so-called ``two-ribbon" flares 
occurring in the biggest and more complex spotted areas. 
Analogously to the H$\alpha$ line, we estimated the peak luminosity for the \ion{He}{i} line as 
$L_{\rm He}\approx9\times10^{28}$ erg\,s$^{-1}$ and the total emitted energy 
as $E_{\rm He}^{\rm tot}\leq10^{34}$ erg. It seems that the flare had a shorter duration in the \ion{He}{i} line, 
because the $EW_{\rm HeI}$ of the last spectrum is at the same level as in the quiescent phase. 

\begin{table}   
\caption[Temperature values and parameters of the subtracted spectra of II~Peg]{Temperature values and parameters 
of the subtracted spectra of II~Peg.}
\label{tab:halpha_ii}
\scriptsize
\begin{center}
\begin{tabular}{ccccc}
\hline
\hline
HJD & Phase & $T_{\rm eff}$ & $\Delta EW_{\rm H\alpha}$ & $EW_{\rm HeI}$\\ 
(+2\,400\,000) & & (K) & (\AA) & (\AA) \\ 
\hline
51\,856.359 & 0.564 & 4546$\pm$ 4& 2.54$\pm$0.11& $~~0.012\pm$0.015\\	 
51\,858.328 & 0.857 & 4478$\pm$29& 3.15$\pm$0.21& $~~0.028\pm$0.038\\	 
51\,859.422 & 0.020 & 4525$\pm$34& 2.83$\pm$0.13& $~~0.019\pm$0.021\\	 
51\,860.348 & 0.157 & 4525$\pm$11& 2.24$\pm$0.12& $~~0.032\pm$0.028\\	 
51\,861.328 & 0.303 & 4573$\pm$ 7& 1.84$\pm$0.11& $~~0.030\pm$0.017\\	 
51\,862.379 & 0.460 & 4570$\pm$25& 2.10$\pm$0.12& $~~0.034\pm$0.018\\	 
51\,863.383 & 0.610 & 4514$\pm$20& 2.62$\pm$0.12& $~~0.047\pm$0.025\\	 
51\,864.488 & 0.773 & 4446$\pm$20& 3.05$\pm$0.13& $~~0.013\pm$0.038\\	 
51\,865.379 & 0.906 & 4453$\pm$10& 5.19$\pm$0.15& $-0.167\pm$0.051\\	  
51\,866.457 & 0.066 & 4495$\pm$23& 3.00$\pm$0.16& $~~0.020\pm$0.027\\	  
\hline
\hline
\end{tabular}
\end{center}
\end{table}
\normalsize

\section{The spot/plage model}
\label{sec:spot}

We have shown in Paper~II that, with a spot model applied to contemporaneous light and temperature curves, it is 
possible to reconstruct the starspots distribution and to remove the degeneracy of solutions regarding spot temperature 
and areas. 
Our spot model assumes two circular active regions whose flux contrast ($F_{\rm sp}/F_{\rm ph}$) can be evaluated 
through the Planck spectral energy distribution (SED), the ATLAS9 \citep{Kuru93} and PHOENIX NextGen \citep{Haus99}
atmosphere models.  

In Paper~II we have evaluated temperature and sizes for the starspots observed on VY~Ari, IM~Peg, and HK~Lac in the fall 
of 2000. We have also shown that both the atmospheric models (ATLAS9 and NextGen) provide values of the spot temperature, 
$T_{\rm sp}$, and area coverage in close agreement, while the black-body assumption for the SED leads to underestimate 
the spot temperature. This result is also in agreement with the findings of \citet{Ama99}. 
Since we have no long-term record of the photospheric temperature, we have assumed the maximum value obtained in our
observing runs as the ``unspotted'' temperature for the modeling. 

In order to analyze the chromospheric rotational modulations, we have extended our spot model, allowing also for bright 
active regions, with the aim of further investigating the degree of spot-plage correlation. 
Given the scatter in the data, two bright spots (plages) are  fully sufficient to reproduce the observed variations. 

A parameter that must be fixed in the model is the flux ratio between plages and surrounding chromosphere 
($F_{\rm plage}/F_{\rm chrom}$). Values of $F_{\rm plage}/F_{\rm chrom}\,\approx\,2$, that can be deduced averaging 
solar plages in H$\alpha$ \citep[e.g., ][]{Sves76,Ayres86}, are too low for modeling the high amplitudes of H$\alpha$ 
emission curves observed for $\lambda$~And and II~Peg. In fact, extremely large plages, covering a significant fraction of
the stellar surface, would be required with such a low flux ratio and they could not reproduce the observed modulations. 
On the other hand, very high values of flux ratio ($F_{\rm plage}/F_{\rm chrom}\,\gtsim 10$) would imply very small plages 
producing top-flattened modulations that are not observed.
So the flux ratio was fixed to values in the range 3--8, that is also typical of the brightest parts of solar plages or 
of flare regions. Note that \citet{Lan00} find $F_{\rm plage}/F_{\rm chrom}\approx 3$ for HR~1099, which has $T_{\rm eff}$
and $\log g$ similar to II~Peg.

The solutions essentially provide the longitude of the plages, giving only rough estimates of their latitude and size. 
We have searched for the best solution by varying the longitudes, latitudes and  radii of the active regions. The radii 
are, however, strongly dependent on the assumed flux contrast $F_{\rm plage}/F_{\rm chrom}$. Thus, only the combined 
information between plage dimensions and flux contrast, i.e. some kind of plage ``luminosity'' in units of the quiet 
chromosphere ($L_{\rm plage}/L_{\rm quiet}$) can be deduced as a meaningful parameter. Note also that we cannot estimate 
the true quiet chromospheric contribution (network), since the H$\alpha$ minimum 
value, $\Delta EW_{\rm quiet}$, could be still affected by a homogeneous distribution of smaller plages.

\subsection{$\lambda$~Andromedae}

From the unspotted temperature and magnitude of $\lambda$~And, namely $T_{\rm ph}=4740$ K and the historical light maximum
$V_{\rm max} = 3\fm70$ \citep{Boyd83}, and the Hipparcos parallax of 38.74$\,\pm\,$0.68 mas, the derived star radius was 
$R = 7.51$ $R_{\odot}$. 
The inclination of the rotation axis with respect to the line of sight derived from this value of star radius, the 
$v\sin i$=6.5 km\,s$^{-1}$ \citep{Dona95}, and the rotation period comes out to be $i\simeq 67\degr$.
\citet{Dona95} found the same value of the star radius and estimated an inclination $i = 60\degr_{-15}^{+30}$. 
Since our $i$ value is the same as \citep{Dona95} within the errors, we adopted $i = 60\degr$ for the spot modeling.

\begin{figure}
\vspace{-.5cm}
\includegraphics[width=9cm]{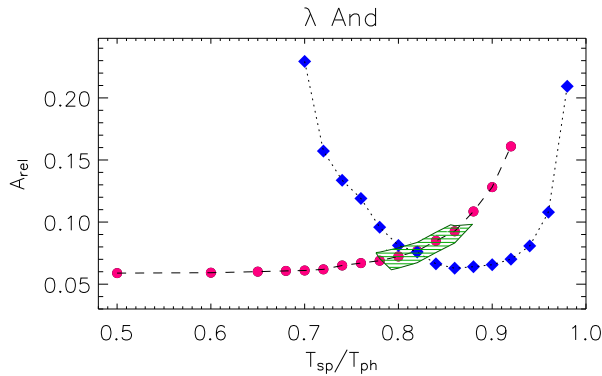}
\caption{Grids of solutions for $\lambda$~And. The filled circles represent the values of spot temperature and area 
for the light-curve solutions, while the diamonds represent the solutions for the temperature curve. 
The hatched area, in each box, is the locus of the allowed solutions accounting for the data uncertainty.}
\label{fig:grids_lam}
\end{figure}

The results of the grids of solutions for the light curve and the temperature curve are displayed in Fig.~\ref{fig:grids_lam}, 
while the spot/plage configuration is given in Table~\ref{tab:parameters_lam} and displayed in Fig.~\ref{fig:lambda_and_tm_v_ew} 
which also shows the synthetic curves superimposed to the data as full lines. In this case $F_{\rm plage}/F_{\rm chrom}$ was 
set to 8. A smaller contrast value would give rise to very big plages with a worse fitting of the H$\alpha$ curve.

\begin{table}  
\caption{Spot/plage configuration for $\lambda$~And$^{a}$. }
\label{tab:parameters_lam}
\scriptsize
\begin{tabular}{crcccc}
\hline
\hline
   Radius    &  Lon.     &   Lat.    &  $T_{\rm sp}/T_{\rm ph}$ & $T_{\rm sp}$ &   $A_{\rm rel}$\\ 
   ($\degr$) & ($\degr$) & ($\degr$) &                                  &      (K)     &                \\ 
\hline
\multicolumn{6}{c}{S{\sc pots}}\\
\hline
26.5   &  343  &  57  & 0.815$^{+0.064}_{-0.036}$ & 3861$^{+304}_{-170}$ & 0.076$^{+0.023}_{-0.014}$ \\
17.1   &   64  &   9  &  & &      \\
\hline
\multicolumn{6}{c}{P{\sc lages}}\\
\hline
22.5   &  360  &  63    &   &  & 0.068 \\
19.8   &   48  &  18    &   &  &  \\
\hline
\hline
\end{tabular}
\begin{flushleft}
$^{a}$ Limb-darkening coefficients $\mu_{\rm V}$=0.797 and $\mu_{6200}$=0.68, $T_{\rm ph}= 4738$\,K, 
$\Delta EW_{\rm quiet}$=0.114 \AA, and $F_{\rm plage}/F_{\rm chrom}=8$ have been used.
\end{flushleft}
\end{table}
\normalsize

For $\lambda$~And, \citet{BoppNoah80}  and \citet{PoeEaton85} found that the asymmetric shape of the light curve requires two 
spots at different longitudes and that these spots are 800--1050 K cooler than the surrounding photosphere, in close
agreement with our findings. The two spots, 
revealed in the images of \citet{Dona95} and covering altogether about 12\% of the total stellar surface, had temperatures 
of 4000$\pm$300\,K, while the photospheric temperature was determined to be 4800$\pm$50\,K, giving a temperature difference 
$\Delta T\simeq800$\,K. They also calculated that the strong global magnetic fields recently detected in RS~CVn systems could 
provide strong enough magnetic braking to explain the observed non-synchronization of $\lambda$~And's rotation. Finally, 
\citet{PadPan99} find $\Delta T = 800\pm30$\,K and $A_{\rm rel}=8.5\pm0.3$ \% setting $T_{\rm ph}=4800$ K and $i=60\degr$. 
The value of $\Delta T\simeq850$ K derived by us is in close agreement with all these previous determinations, but 
\citet{Oneal98b} find cooler spots ($T_{\rm sp}\simeq3650$\,K) by using the TiO bands.

\subsection{II~Pegasi}

For II~Peg we adopted unspotted magnitude and temperature values of $V_{\rm max} = 6\fm9$ \citet{Berdy98b}  and 
$T_{\rm ph}=4573$ K (present work). From the unspotted and de-reddened $V$ magnitude and the Hipparcos parallax of
23.62 mas, a star radius of 2.76 $R_{\odot}$ is derived with the Barnes-Evans relation.
From the radius, the rotational period of $6\fd724$, and the $v\sin i$ value of 22.6 km\,s$^{-1}$, an inclination of 
$60\degr_{-10}^{+30}$ has been deduced. The mass we derived from the evolutionary tracks of \citet{Gira00}  for 
the primary component is 0.9 $M_{\odot}$. 
\citet{Berdy98a} derived a mass of $0.8\pm0.1$ $M_{\odot}$ and a radius of $3.4\pm0.2$ $R_{\odot}$ ($\log g\simeq 3.2$) for 
the primary component and a mass of $0.4\pm0.1\,M_{\odot}$ for the unseen secondary component, whose spectral type has been 
estimated to be M0--3V.

\begin{figure}[h]
\vspace{-1.5cm}
  \begin{center}
  \includegraphics[width=9cm]{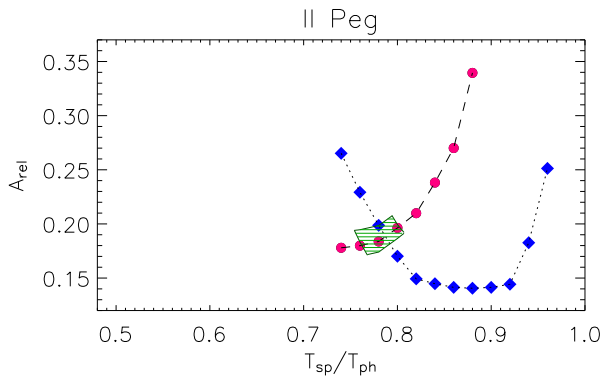}
  \end{center}
\caption{Grids of solutions for II~Peg.  The filled circles represent the values of spot temperature and area 
for the light-curve solutions, while the diamonds represent the solutions for the temperature curve. 
The hatched area, in each box, is the locus of the allowed solutions accounting for the data uncertainty.}
\label{fig:Teff_aree_ii_err_K}
\end{figure}

The result of the intersection of the grids of solutions for the light curve and the temperature curve 
(Fig.~\ref{fig:Teff_aree_ii_err_K}) provides a $T_{\rm sp}/T_{\rm ph}$ of 0.787, i.e a $T_{\rm sp}\simeq3600$\,K.
However,  the spot model with NextGen SED has been also applied and similar values to 
those presented in Table~\ref{tab:parameters_ii} have been found. 

To reproduce the $\Delta EW_{\rm H\alpha}$ curve a relative flux $F_{\rm plage}/F_{\rm chrom}> 2$ was needed. 
A value of $F_{\rm plage}/F_{\rm chrom}= 4$ gives a satisfactory fit of the data, though, as previously outlined, 
this parameter cannot be constrained by the modeling of the H$\alpha$ curve.
However, the analysis carried out by \citet{Busa99} and \citet{Lan00} on HR~1099, which has $T_{\rm eff}$ and 
$\log g$ similar to II~Peg, points to a contrast of $F_{\rm plage}/F_{\rm chrom}\approx 3$.
Such analysis, based on the modeling of the \ion{Mg}{ii} h \& k and H$\alpha$ profiles, gives more
constraints on the the flux contrast and the AR area.  

The configuration of the active regions is represented in Fig.~\ref{fig:iipeg_tm_v_ew} and the active region parameters are 
listed in Table~\ref{tab:parameters_ii}.

The first work on spot modeling for II~Peg was made by \citet{BoppNoah80}, 
who showed that, for a satisfactory modeling of their asymmetric light curve, two cool spots were sufficient. 
\citet{Vogt81a} and \citet{HueRam87} made a 
quantitative study of the effect of spots in the TiO bands and found that a substantial fraction of the photosphere 
must be spotted (with a coverage of 35--40\%). Then, \citet{Vogt81b} found that the light and color curves could 
be reproduced with a cool spot having an effective temperature of 3400$\pm$100 K and covering about 37\% of one 
hemisphere on the star. 
In subsequent years, several other authors have produced spot models for several 
datasets obtained in different epochs and derived spot temperatures. Among these, the more relevant works are those 
of \citet{NatRam81},
\citet{PoeEaton85}, 
\citet{Rodo86},     
\citet{ByrMar87},   
and \citet{Boyd87}. 

\citet{Dona97}, by means of the Zeeman-Doppler Imaging technique, which takes advantage of the Doppler effect 
in rapidly rotating stars to separate in wavelength the disk-integrated polarized profiles of magnetic lines, 
detected clear signatures of magnetic field with possible concentrations near the longitudes of the spots observed 
by \citet{Henry95}. Moreover, it was shown for this star that the changes in the light curve of II~Peg are consistent 
with re-arrangements of the spot distribution over the stellar surface. \citet{Neff95}, from TiO molecular 
bands, found that cool starspots ($T_{\rm sp}$ = 3500$\pm$200 K) are always visible, with a fractional projected 
coverage of the visible hemisphere varying from 54\% to 64\% as the star rotates. 
\citet{Oneal97} 
detected excess of OH absorption due to cool spots on the surface of II~Peg and found a spot filling factor of 
35--48\%, consistent with the minimum value of 40\% found by \citet{Mari99}. \citet{Hatzes95} revealed 
polar or high-latitude spots and several equatorial spots with a total coverage of about 15\%. \citet{Oneal98}  
reported the first spectroscopic evidence for the multiple spot temperature for this star. From TiO molecular 
band observations, they found that spot temperature varied between 3350 to 3550 K during the epoch of 
September 1996 to October 1996, whereas the starspot filling factor was constant (about 55\%). Finally, \citet{Berdy98b} 
found, from their surface images, that the high-latitude spots were the major contribution to the 
photospheric activity of II~Peg. \citet{PadPan99}  
obtained $\Delta T = 740\pm45$ K and $A_{\rm rel}=15.7$ setting $T_{\rm ph}=4300$ K and $i=34\degr$. 
Recently, \citet{Gu03}, from the Doppler imaging analysis, have 
found for II~Peg changes in spot distribution, including the position, intensity and size of the spots occurring in 
1999-2001. In particular, they find in February 2000 a large high-latitude spot and a small weak low-latitude 
spot, i.e. the same spot configuration found in the present work for the observing season of November 2000.

\begin{table}[h]   
\caption{Spot/plage configuration for II~Peg$^{a}$. }
\label{tab:parameters_ii}
\scriptsize
\begin{tabular}{crcccc}
\hline
\hline
   Radius    &  Lon.     &   Lat.    &  $T_{\rm sp}/T_{\rm ph}$ & $T_{\rm sp}$ &   $A_{\rm rel}$\\ 
   ($\degr$) & ($\degr$) & ($\degr$) &                                  &      (K)     &                \\ 
\hline
\multicolumn{6}{c}{S{\sc pots}}\\
\hline
 47.8   &  320  &  60  & 0.787$^{+0.020}_{-0.041}$ & 3599$^{+91}_{-188}$ K & 0.189$^{+0.019}_{-0.018}$ \\
 16.8   &  202  &   8  &  & &      \\
\hline
\multicolumn{6}{c}{P{\sc lages}}\\
\hline
 35.9   &  322  &  60    &	&  & 0.132 \\
 18.8   &  206  &   8    &	&  &  \\
\hline
\hline
\end{tabular}
\begin{flushleft}
$^{a}$ Limb-darkening coefficients $\mu_{\rm V}$=0.836 and $\mu_{6200}$=0.74, $T_{\rm ph}= 4573$\,K, 
$\Delta EW_{\rm quiet}$=1.728 \AA, and $F_{\rm plage}/F_{\rm chrom}=4$ have been used.
\end{flushleft}
\end{table}

\section{Discussion}

The solar paradigm can serve as a guide for the interpretation of the behaviors of stellar ARs at
different atmospheric layers. 
In the Sun,  it is found that spots are always spatially 
associated with faculae, although the reverse is not always true. Indeed, faculae generally form before a spot
or a spot group appears and survive for some time after the disappearance of the spots. Moreover, high-latitude
small faculae or plages not associated to spots are normally observed on the Sun. The ARs trace
magnetic flux tube emersion, that is influenced by several effects, such as the Coriolis force, magnetic tension,  
and convective motions.
Therefore, studies of the topology, motions, and orientations of ARs at different levels (spots, plages) can provide 
information about the physical processes occurring below the solar/stellar surface.

Some papers dealt with the behavior of facular to sunspot areas along the solar activity cycle 
\citep[e.g., ][]{Chapman97,Foukal98}. In particular, \citet{Foukal98} found that the area ratio of
faculae to spots decreases at increasing activity levels both using white-light faculae and \ion{Ca}{ii} K
chromospheric plages and explains this as a consequence of the different dependencies of plage and spot
lifetimes upon their emergent magnetic flux. Thus, the sub-photospheric field properties are believed to be
more important to determine this ratio, rather than photospheric field diffusion. 
Moreover, this shift toward dark photospheric structures for high activity levels can explain the high
variation amplitudes observed in late-type stars more active than the Sun.

Some authors  found that for moderately active stars the variability at optical wavelength is strongly
influenced by faculae \citep[e.g., ][]{Radick98,Mirto03}. However, the facular contribution has been proposed
also for very active stars by other authors to account for the UV flux excess of active stars compared
to non-active ones \citep{Ama03} and to explain the blueing of the colors when the stars get fainter 
\citep[e.g., ][]{Aarum03,Messina06}. The presence of faculae/plages around spots seems to be an ubiquitous
phenomenon \citep[e.g., ][]{Fra97,Fra98,Cata00,Bia06,Bia07} and the non-detections could be due to contrast reasons.

Despite the uncertainty in deriving the radii of the plages from our H$\alpha$ variation curves, we would
like to outline that the plage area is larger than that of the underlying spot for the smaller AR, while the reverse 
is true (the spot bigger than the plage) for the larger AR. This holds true both for $\lambda$~And and II~Peg and
is in line with the findings of \citet{Foukal98} for solar ARs.

As regards the degree of spatial correlation of spots and plages in active stars, guidelines can be represented by 
disk-integrated observation of the Sun.  \citet{Cata98} report on a strong rotational modulation of the solar irradiance 
in the \ion{C}{ii} $\lambda$\,1335\,\AA ~chromospheric line from UARS SOLSTICE experiment.
They ascribe to chromospheric plages this modulation and show that it is highly correlated with the sunspot number. 
The average position of spots appears to alternately lead and lag the centroid of plages by about 30--40$\degr$, 
at maximum, with a possible period of 270 days. 
Some indication of small longitude shifts between H$\alpha$ plages and spots within this range has been found for very 
active stars by \citet{Cata00}.
We find that the plage longitudes in II~Peg are nearly the same as those of the underlying spots. For $\lambda$~And, instead,
longitude shifts of about $+17\degr$ and $-16\degr$ for the two ARs, that we consider as marginally significant due to the data scatter and phase
coverage, come out from the model solution.

The investigation of possible dependencies of spot parameters like temperature and filling factor on global stellar
parameters like effective temperature, gravity, activity level (rotation rate, differential rotation, etc.) is of 
great importance to better understand the physical mechanisms at work on the formation and evolution of ARs.

Recently, a correlation of temperature difference between the quiet photosphere and spots, $\Delta T$, with the 
effective temperature has been found by \citet{Berdy05}. She found that, on average, $\Delta T$ is larger for the
hotter stars, with values of nearly 2000\,K for stars with $T_{\rm eff}\approx6000$\,K and falling down to about 200\,K for 
M4 stars. This behavior is displayed both by giant and main-sequence stars.

It is interesting to investigate the role of the surface gravity on $\Delta T$ by selecting stars of nearly the same 
temperature.
To further investigate this issue, we have also used the values of $\Delta T$ derived by us in Paper~II for the three active 
giant/subgiant stars with the LDR method. In addition, we have also considered the values of  $\Delta T$=1325\,K and 1030\,K
obtained, with a typical error of $\pm 150$\,K, by \citet{Oneal01,Oneal04} with the method of TiO bands for the two 
main sequence stars V833~Tau and EQ~Vir, which have $T_{\rm eff}$ similar to the star investigated by us and 
$\log g\simeq 4.5$. 
The data of these seven active stars suggest an increasing trend of $\Delta T$ with gravity, as already proposed by 
\citet{Oneal96}, that can be explained by the balance of magnetic and gas pressure in the flux tubes of active regions. 
However, more data are needed to further investigate this correlation.

\section{Conclusion}
We have presented a study of the surface inhomogeneities at both  photospheric and chromospheric levels
based on a contemporaneous spectroscopic and photometric monitoring of the two active RS~CVn stars  
$\lambda$ Andromedae and II~Pegasi.

From the same data set of medium-resolution optical spectra, we have obtained information
about the chromospheric and photospheric surface inhomogeneities by using the H$\alpha$ emission and the 
photospheric temperature (from line depth ratios), respectively. 
Additional information coming from the light curves, together with the temperature modulations, allowed us to 
disentangle the effects of starspot temperature and area and deduce these parameters in a unique way.
A very important indication from this work is that the the starspots of $\lambda$~And are considerably smaller and
warmer than those of II~Peg, notwithstanding the nearly equal photospheric temperature.
At present we cannot say if this is due to the different gravity of the two active stars, being $\lambda$~And a  
giant with $\log g\simeq$\,2.5 and II~Peg a subgiant ($\log g\simeq$\,3.2), or if it is simply the 
effect of the higher activity level of II~Peg compared to  $\lambda$~And.
More active stars with different gravity and activity level must be investigated to settle this point.
 However, by using the values of temperature difference between photosphere and spot, $\Delta T$, for 
other stars from our previous works and from the literature, we find  an increasing trend of $\Delta T$ 
versus $\log g$ that  could be explained by the magnetostatic equilibrium between gas and magnetic pressure. 

We have found, for all the stars observed, a tight anti-correlation between the H$\alpha$ emission and the 
photospheric temperature modulations, that indicates a close spatial association between photospheric spots 
and chromospheric plages.  The largest longitude shifts between plages and spots of about $\pm 16\degr$,
have been found for $\lambda$~And. Moreover, the area ratio of plages to spots seems to decrease when the spots 
get bigger.

Furthermore, in II~Peg a strong flare affecting both the H$\alpha$ and \ion{He}{i} lines has been observed.
The energy losses in these lines have been also evaluated.
A possible flare event, with a much smaller energy budget, seems to have occurred in  $\lambda$~And.
In the present work we have shown the great power of a coordinated photometric and spectroscopic monitoring of
active stars for the study of the main properties of their active regions.

\begin{acknowledgements}

We are grateful to an anonymous referee for helpful comments and suggestions.
This work has been supported by the Italian {\em Ministero dell'Istruzione, Universit\`a e  Ricerca} (MIUR) and by the 
{\em Regione Sicilia} which are gratefully acknowledged. We also thank the Scientific and Technological Research
Council of Turkey (T\"{U}B\.{I}TAK).
This research has made use of SIMBAD and VIZIER databases, operated at CDS, Strasbourg, France.

\end{acknowledgements}

\bibliographystyle{aa}

\end{document}